\documentclass{article}
\usepackage{multirow}
\usepackage{tikz}
\usepackage{arxiv}
\usepackage[utf8]{inputenc} % allow utf-8 input
\usepackage[T1]{fontenc}    % use 8-bit T1 fonts
\usepackage{hyperref}       % hyperlinks
\usepackage{url}            % simple URL typesetting
\usepackage{booktabs}       % professional-quality tables
\usepackage{amsfonts}       % blackboard math symbols
\usepackage{nicefrac}       % compact symbols for 1/2, etc.
\usepackage{microtype}      % microtypography
\usepackage{lipsum}
\usepackage{graphicx}
\usepackage{amsmath}
\usepackage{amssymb}
\usepackage{graphicx}
\usepackage{natbib}
\usepackage{float}
\usepackage{tikz}

\usepackage{multirow}
\usepackage{booktabs}
\usepackage{makecell}
\usetikzlibrary{shapes.geometric, arrows.meta, positioning}

\tikzstyle{box} = [rectangle, rounded corners, minimum width=4.5cm, minimum height=1.6cm,text centered, draw=black, fill=gray!10]
\tikzstyle{arrow} = [thick,->,>=stealth]
\graphicspath{ {./images/} }

\title{From Pen to Palette: Mathematical Analysis of van Gogh's Psychological Trajectory}

\author{
 Anna Singley \\
  Department of Mathematics\\
  University of Portland\\
  Portland, OR 97203 \\
  \texttt{singley26@up.edu} \\
   \And
 Eli Goldwyn \\
  Department of Mathematics\\
  University of Portland\\
  Portland, OR 97203 \\
  \texttt{goldwyn@up.edu} \\
  \And
 Mark Pitzer \\
  Department of Psychology \\
  University of Portland\\
  Portland, OR 97203 \\
  \texttt{pitzer@up.edu} \\
  \And
 Jessica Logue \\
  Department of Philosophy\\
  University of Portland\\
  Portland, OR 97203 \\
  \texttt{logue@up.edu} \\
  }
\begin{document}
\maketitle
\begin{abstract}
Vincent van Gogh's prolific artistic output and personal correspondence offer a unique window into the psychological trajectory of a 19th-century artist. This study applies computational methods to analyze emotional patterns across van Gogh's final decade through sentiment analysis of over 700 letter segments and fractal dimension analysis of 569 paintings. Using critical slowing down (CSD) theory from dynamical systems, we examine whether van Gogh's psychological system exhibited mathematical signatures of approaching tipping points. Our analysis reveals a statistically significant correlation between visual complexity and death sentiment ($r = 0.193, p < 0.001, n = 524$), alongside evidence for CSD patterns in multiple psychological dimensions before major crises. The 1888 ear incident shows coordinated variance spikes across acquired capability, desire for death, and visual complexity preceding the event, followed by stabilization before the crisis. These findings demonstrate that mathematical approaches can detect psychological transition signatures in historical figures while revealing the complex multi-modal nature of psychological systems.

\end{abstract}

% keywords can be removed

\section{Introduction}

Vincent van Gogh was a famed post-impressionist painter known almost as well for his tragic life story as for his art. van Gogh's final decade (1880-1890) represents one of the most documented psychological trajectories in art history, culminating in his death in July 1890. The van Gogh Museum's digitization of his extensive correspondence with brother Theo, combined with his prolific artistic output during periods of psychological distress, enables computational analysis of historical psychological dynamics at unprecedented temporal resolution \citep{museum}.

While qualitative analyses have long examined van Gogh's materials for evidence of psychological patterns \citep{bekker2009color}, this study applies computational methods to identify and quantify signals of emotional instability across his final years. We analyze van Gogh's psychological state through two distinct yet complementary modalities: sentiment expressed in over 700 letter segments and geometric complexity measured through the fractal dimension of 569 paintings. 

%Specifically, we measure seven sentiments commonly linked to self-harm by using a supervised sentiment classification and we compute the fractal dimension to measure the geometric complexity of the painting.

Previous computational work has examined multifractal texture qualities of works from different periods revealing greater regularity in Paris-period paintings compared to later works \citep{abry2013van}, and studies of luminance patterns in his paintings found that stormy compositions exhibit mathematical structures closely resembling real turbulent flow \citep{aragon2008turbulent}. However, these studies analyzed subsets of paintings or specific visual features rather than systematic measurements across his entire oeuvre. 

Our approach draws from nonlinear dynamical systems theory, specifically the concept of critical slowing down (CSD), in which increases in variance, autocorrelation, and higher-order statistical moments signal a system's proximity to phase transitions \citep{dakos2012robustness}. Originally developed in ecology and climate science to help predict ecosystem collapse \citep{dakos:2014} and climate tipping points \citep{scheffer}, CSD theory has recently shown promise in predicting psychological transitions, including mood disorder episodes and suicide risk \citep{singley2025stochastic}. We hypothesize that van Gogh's psychological system may exhibit the statistical precursors of tipping points detectable through both textual and visual expression channels.

This framework is further supported by the medical hypothesis that van Gogh may have suffered from acute intermittent porphyria (AIP). AIP is a hereditary metabolic disorder, is highly episodic in nature, and is characterized by attacks that can leave a patient severely incapacitated followed by periods of complete symptomatic remission \citep{loftus1991vincent} The resulting high variance symptom severity may have contributed to the psychological instability associated with self-harm.

By modeling van Gogh's creative and emotional fluctuations as a dynamical system, we integrate digital humanities, computational art history, and historical psychology. This interdisciplinary approach offers both understanding of van Gogh's psychological trajectory and a replicable methodology for detecting  transitions in historical figures across diverse media.

\section{Methods}\label{sec2}

\subsection{Data Sources}
The letter sentiment data comprises 719 text segments extracted from van Gogh's correspondence, representing over 900 individual letters sent between 1872-1890 \citep{van1963letters}. Letters containing only sketches were not included in our sentiment analysis work. We analyze 569 paintings spanning 1881-1890 with documented completion dates, sourced from digitized museum collections and scholarly catalogs \citep{museum}. Temporal alignment used a 30-day proximity window, yielding 524 matched data points with mean temporal separation of 2.1 days between painting and letter data.

\subsection{Letter Analysis}

We analyze van Gogh's emotional trajectory using a supervised sentiment classification pipeline applied to letters sent between 1872 and 1890. Each letter was segmented into paragraphs by line break, which served as discrete units for sentiment evaluation. This choice allowed a letter to contain varying levels of sentiment or opposing valences within the same correspondence.

The sentiment classifier is a support vector machine (SVM) with a radial basis function (RBF) kernel, trained on a labeled social media dataset consisting of over 100,000 posts categorized using psychological indicators from established clinical frameworks \citep{stitson1996theory} \citep{thurnhofer2020radial}. Sentiments are selected based on both psychological theory and relevance to van Gogh's documented experiences are: desire for death, acquired capability, social isolation, anger, drug/alcohol use, stress, and social connection. These categories align with interpersonal theory of suicide, which identifies thwarted belongingness and perceived burdensomeness as core risk pathways \citep{van2010interpersonal}.

Text was encoded using count vectorization, and the SVM model outputs were transformed into continuous sentiment scores between 0 and 1 using Platt scaling, which maps the distance from the canonical hyperplane to a probability-like confidence score \citep{boken2021appropriateness}. This enables us to compare not only valence, but strength of sentiment across different emotional categories. Examples of this methodology with sentiment scores appear in Table \ref{tab:sentiment}. Poetry excerpts are shown for illustration, as the van Gogh correspondence is too extensive for effective tabular presentation. Each paragraph received a score for each of the sentiment categories. 

\begin{table*}[t]
\centering
\small
\begin{tabular}{|c|l|l|p{8cm}|c|c|c|}
\hline
\textbf{id} & \textbf{author} & \textbf{Title} & \textbf{content} & \textbf{anger} & \textbf{happiness} \\
\hline
1 & Anne Sexton & 45 Mercy Street & Pull the shades down—I don't care! Bolt the door mercy erase the number rip down the street sign what can it matter what can it matter to this cheapskate who wants to own the past that went out on a dead ship and left me only with paper? & 0.923669 & 0.2730928  \\
\hline
2 & Shakespeare & Romeo And Juliet & The dashing rocks thy sea-sick weary bark! Here's to my love! O true apothecary! Thy drugs are quick. Thus with a kiss I die. & 0.993142 & 0.034483 \\
\hline
3 & Sylvia Plath & Edge & The moon has nothing to be sad about, Staring from her hood of bone. She is used to this sort of thing. Her blacks crackle and drag. & 0.87426 & 0.03506799 \\
\hline
\end{tabular}
\caption{Sentiment analysis of literary excerpts.}
\label{tab:sentiment}
\end{table*}

\subsection{Painting Analysis}

In parallel with our analysis of van Gogh's letters, we apply computational image analysis techniques to quantify changes in the formal properties of his paintings. Our objective is to assess whether fluctuations in visual complexity — measured through fractal geometry — could serve as a correlate to emotional instability during critical periods in van Gogh's life.

\begin{figure}
    \centering
    \includegraphics[width=0.9\linewidth]{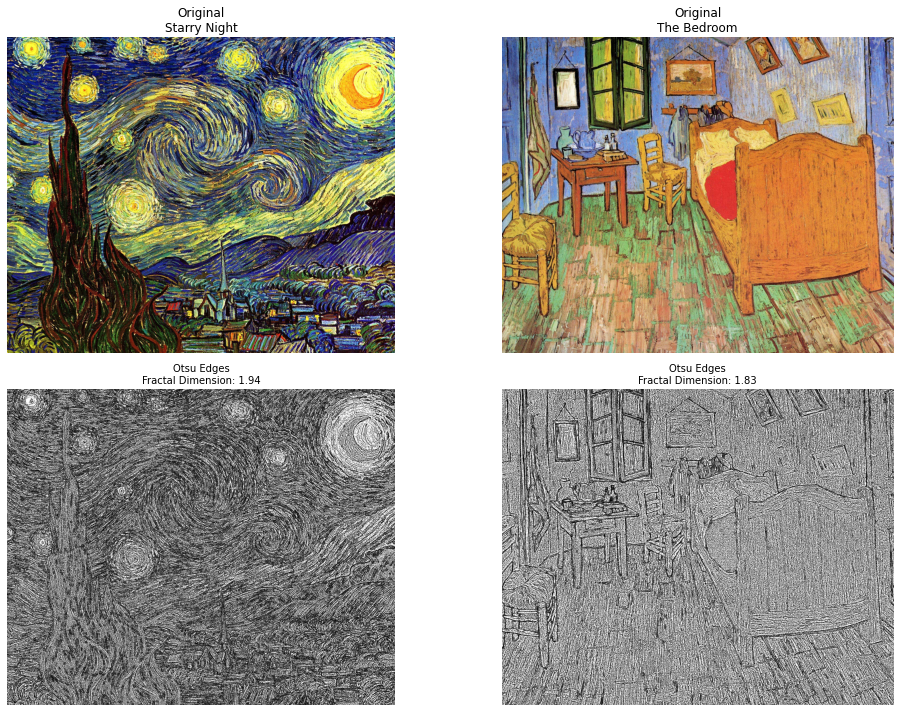}
    \caption{The detected brushstrokes of two paintings with their corresponding Minkowski dimension. Starry Night, with a fractal dimension of 1.93 and painted in June 1889, and The Bedroom, with a fractal dimension of 1.82 and painted in October 1888.}
    \label{fig:brushstroke_original}
\end{figure}

We collected high-resolution scans of paintings created between 1880 and 1890, aligning temporally with the letter corpus. To isolate features corresponding to brushwork, each image is converted to grayscale and preprocessed using a Difference of Gaussians (DoG) filter \citep{poon1994real}. This technique accentuates edges at multiple spatial scales, effectively highlighting stroke contours. To further enhance continuity in brushstroke structure, we applied line integral convolution (LIC), followed by Otsu's thresholding to binarize the image, see Figure \ref{fig:brushstroke_original}. This pipeline ensured that visual noise was minimized while preserving painterly texture.

\begin{figure}
    \centering
    \includegraphics[width=1\linewidth]{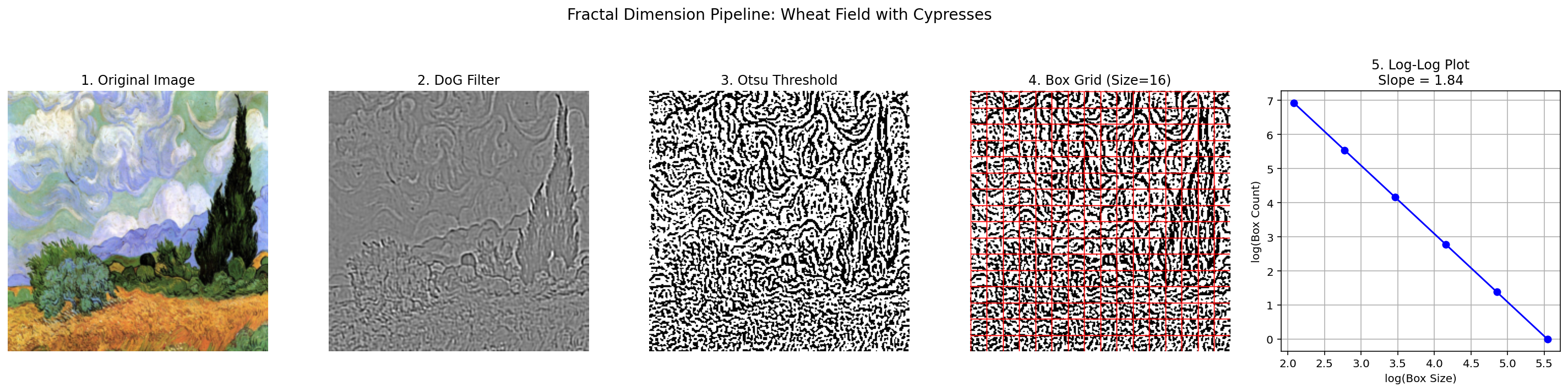}
    \caption{An example of the image processing pipeline employed to obtain the fractal dimension of brushwork in van Gogh paintings.}
    \label{fig:comprehensive_original}
\end{figure}

The processed binary images were then analyzed using a Minkowski–Bouligand box-counting algorithm \citep{falconer:2013} to calculate fractal dimension, see Figure \ref{fig:comprehensive_original}. This metric estimates the scaling complexity of an image and has been used in prior art-historical studies to differentiate between styles, periods, and affective tone \citep{mcdonough2024fractal}. Figure \ref{fig:violin_location} shows the fractal dimension of van Gogh's paintings based on where he lived at the time while Figure \ref{fig:temporal_dist} is sorted based on type of work. In this context, higher fractal dimension indicates denser, more irregular visual structures, which may reflect elevated emotional or cognitive arousal.

\subsection{Critical Slowing Down Analysis}
For both sentiment and visual metrics, we compute rolling statistics using systematically varied window sizes (5-30 data points) to identify optimal temporal resolution for CSD detection. We find an optimal window of one month. Within each of these monthly windows, we compute sentiments and the fractal dimension, along with higher order moments including lag-1 autocorrelation and variance. These higher-order statistical moments are grounded in the theory of critical slowing down and are useful indicators of affective volatility preceding psychological tipping points \citep{nazarimehr2020critical}. We analyze three distinct temporal phases around major biographical events: baseline period (4-6 months before), a theorized critical slowing down phase (2-4 months before), and theorized stabilization phase (0-2 months before).

\section{Results}

\subsection{Cross-Modal Correlations and Temporal Patterns}

Statistical analysis reveals significant correlations between visual complexity (fractal dimension) and multiple psychological dimensions across 524 temporally matched observations with mean separation of 2.1 days between painting and letter data. The strongest associations were observed for stress ($r = 0.115$, $p = 0.008$), substance use ($r = 0.108$, $p = 0.013$), and anger ($r = -0.105$, $p = 0.017$), with acquired capability showing a weaker but significant correlation ($r = 0.096$, $p = 0.028$). The negative correlation between anger and visual complexity suggests that periods of psychological distress characterized by anger corresponded to decreased fractal complexity in artistic output, potentially reflecting cognitive narrowing during emotional dysregulation. Most notably, desire for death demonstrated a significant positive correlation with visual complexity ($r = 0.193$, $p < 0.001$), suggesting that periods of elevated death cognition corresponded to increased fractal complexity in van Gogh's brushwork.

Correlation coefficients represent Pearson's $r$ computed between daily sentiment scores and fractal dimension values after resampling both series to a common daily grid via linear interpolation. Statistical significance was assessed using a two-tailed $t$-test under the null hypothesis of zero correlation, with $p$-values computed from the $t$-statistic $t = r\sqrt{(n-2)/(1-r^2)}$ with $n-2$ degrees of freedom, where $n$ is the number of paired observations.

\begin{figure}
    \centering
    \includegraphics[width=1\linewidth]{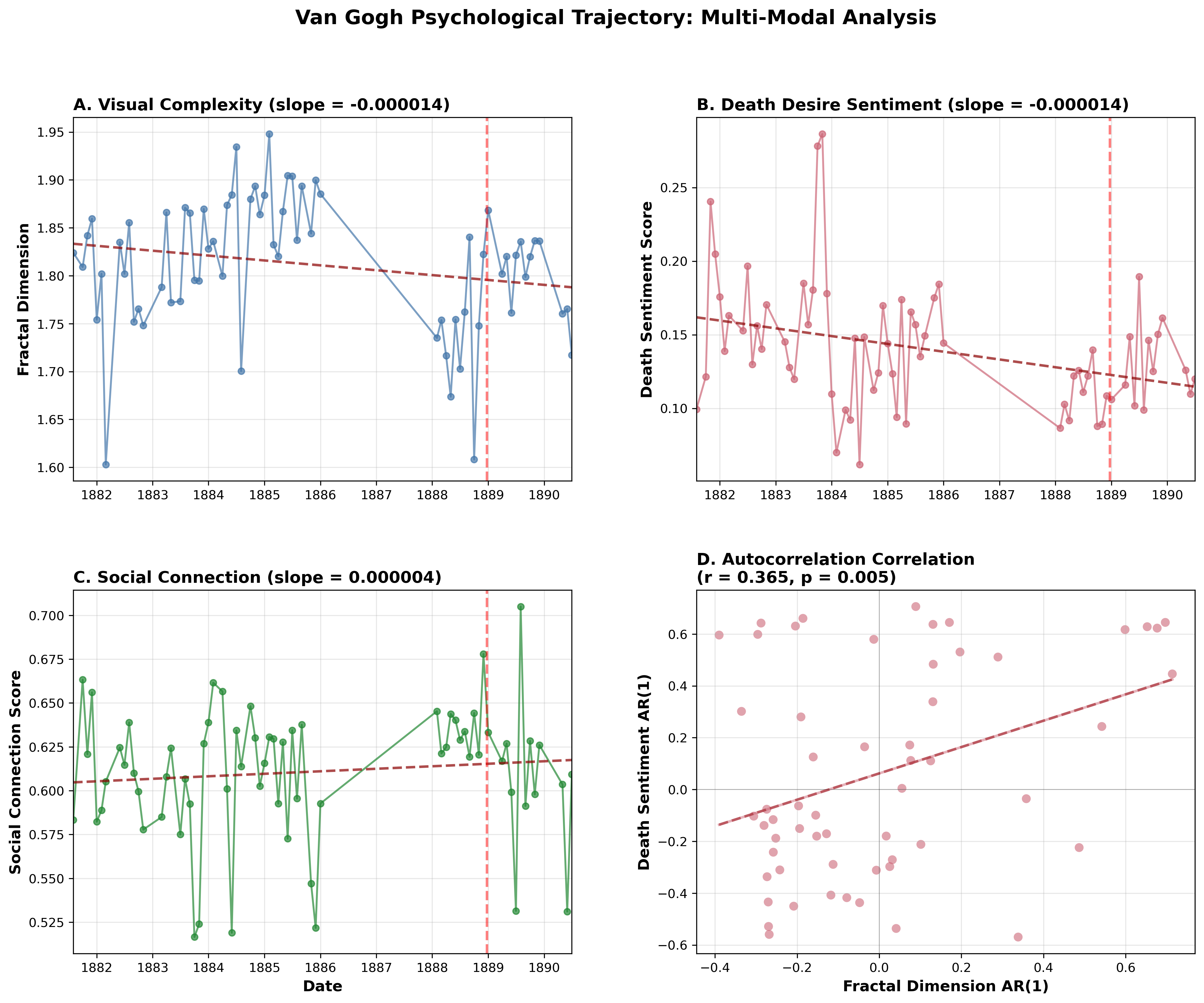}
    \caption{A series of temporal analysis plots of van Gogh's psychological trajectory. (A) Visual complexity (fractal dimension) over time showing a decreasing trend. (B) Temporal trends in the variance of Desire for Death over the timespan of letters. (C) Temporal trends in the variance of Social Connection over the timespan of letters. (D) Correlation of fractal dimension with desire for death in matching time-windows.}
    \label{fig:comprehensive_temporal}
\end{figure}

Analysis of van Gogh's creative output reveals distinct temporal trajectories across psychological and artistic dimensions over his final decade (Figure~\ref{fig:comprehensive_temporal}). Visual complexity shows a statistically significant decreasing trend over time (slope $= -1.4 \times 10^{-5}$ per day, $p < 0.001$, Figure~\ref{fig:comprehensive_temporal}A), indicating progressive simplification of brushwork structure from 1881 to 1890. Desire for death variance exhibits substantial fluctuation across van Gogh's final years with notable increases during documented crisis periods (Figure~\ref{fig:comprehensive_temporal}B), while social connection increased over time (slope $= 4.0 \times 10^{-6}$, $p < 0.001$, Figure~\ref{fig:comprehensive_temporal}C), though this trend masks substantial short-term volatility during crises. The correlation between fractal dimension and desire for death varies across time windows (Figure~\ref{fig:comprehensive_temporal}D), with periods of strong positive correlation interspersed with periods of weak or negative correlation, suggesting that coupling between visual expression and death cognition fluctuated with van Gogh's psychological state.

The fractal dimension of van Gogh's work exhibits subtle variation across geographic locations (Figure~\ref{fig:violin_location}), with Paris-period paintings (1886--1888) demonstrating the highest median fractal dimension ($\text{median} = 1.87$, $\text{IQR} = 1.84\text{--}1.91$), Arles showing intermediate complexity ($\text{median} = 1.85$, $\text{IQR} = 1.81\text{--}1.89$), and Saint-Rémy exhibiting greater variability ($\text{median} = 1.84$, $\text{IQR} = 1.80\text{--}1.88$) despite lower median values. Analysis by work category (Figure~\ref{fig:temporal_dist}) reveals that self-portraits exhibit the highest fractal complexity ($\text{median} = 1.89$, $\text{IQR} = 1.86\text{--}1.92$), landscapes show intermediate complexity ($\text{median} = 1.85$, $\text{IQR} = 1.82\text{--}1.88$), and still lifes demonstrate the lowest median fractal dimension ($\text{median} = 1.83$, $\text{IQR} = 1.80\text{--}1.86$).

\subsection{Critical Slowing Down Before the December 1888 Ear Incident}

\begin{figure}[ht]
    \centering
    \includegraphics[width=\linewidth]{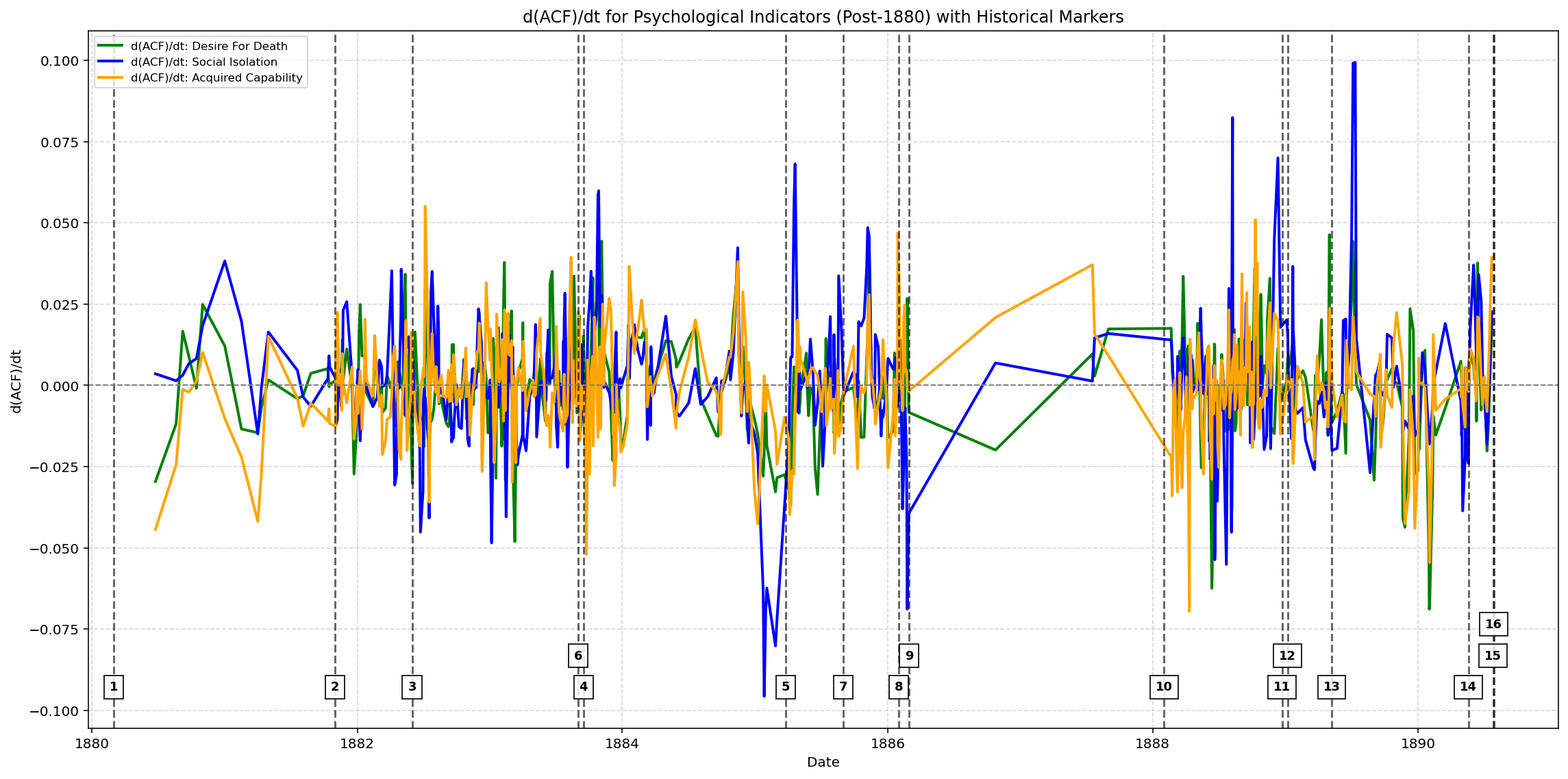}
    \caption{Historical analysis of psychological indicators aligned with major biographical events, demonstrating the temporal relationship between mathematical signatures and documented psychological transitions. The vertical axis is change in autocorrelation.}
    \label{fig:historical}
\end{figure}

The ear-severing incident of December 23, 1888 provides the clearest evidence for critical slowing down signatures in van Gogh's psychological trajectory. Multiple psychological indicators show coordinated patterns of instability beginning approximately 4--6 months before the crisis, providing prospective early warning signals that preceded the documented behavioral emergency. Figure~\ref{fig:historical} displays the temporal evolution of these indicators throughout van Gogh's final decade, with the ear incident marked as event 11.

Examining the period from June through December 1888, multiple psychological dimensions exhibit characteristic CSD patterns visible in Figure~\ref{fig:historical}. The figure shows that acquired capability (orange trace), desire for death (yellow trace), visual complexity (blue trace), and social dimensions (green and purple traces) all show elevated volatility during the months preceding the ear incident. The critical observation is not merely that individual dimensions showed CSD signatures, but that multiple dimensions exhibited coordinated increases during the same 4--6 month window. If these were independent random fluctuations, we would expect temporal decorrelation between dimensions. Instead, the temporal alignment suggests shared underlying dynamics consistent with theoretical models of psychological systems as coupled networks where perturbations propagate across domains.

\begin{table*}
\centering
\caption{Key Historical Events Aligned with van Gogh's Psychological Indicators}
\begin{tabular}{cllp{7.8cm}}
\toprule
\# & Date       & Event                         & Description \\
\midrule
1  & 1880-03-01 & Returned to Cuesmes           & Returned to coal-mining district after family pressure; began sketching workers \\
2  & 1881-11-01 & Hand-in-flame incident        & Self-injury during emotional crisis following rejection by cousin Kee Vos-Stricker \\
3  & 1882-06-01 & Hospitalization for gonorrhoea& Treated in hospital; likely period of stress and illness \\
4  & 1883-09-16 & Margot Begemann suicide attempt& Margot overdosed after family's disapproval of relationship \\
5  & 1885-03-26 & Death of van Gogh's father    & Theodorus van Gogh's sudden death; destabilizing familial event \\
6  & 1883-09-01 & Moved to Nuenen               & Began focused two-year artistic period after relocating from Drenthe \\
7  & 1885-09-01 & Accusation of assault in Nuenen & Accused of impregnating a model; local clergy forbade townspeople from posing \\
8  & 1886-02-01 & Hospitalized in Antwerp           & Possibly due to exhaustion, alcoholism, or untreated illness \\
9  & 1886-03-01 & Moved to Paris                & Joined Theo; entered vibrant Parisian art world; major stylistic transition \\
10 & 1888-02-01 & Departed Paris for Arles      & Sought solitude and sunlight to found artist colony; productivity increased \\
11 & 1888-12-23 & Severed ear incident          & Psychological crisis culminated in self-mutilation; hospitalization \\
12 & 1889-01-07 & Return to Yellow House        & Brief return after hospitalization, still experiencing hallucinations \\
13 & 1889-05-08 & Entered asylum at Saint-Rémy & Voluntarily committed following public petition; painted famous works \\
14 & 1890-05-20 & Moved to Auvers-sur-Oise      & Final period under care of Dr. Gachet; intensified artistic output \\
15 & 1890-07-27 & Suicide attempt               & Shot himself in wheat field; suffered infection and decline \\
16 & 1890-07-29 & Death                         & Died in Theo's presence; last words reportedly "The sadness will last forever" \\
\bottomrule
\end{tabular}
\label{tab:vangogh_events}
\end{table*}

\begin{figure}
    \centering
    \includegraphics[width=1\linewidth]{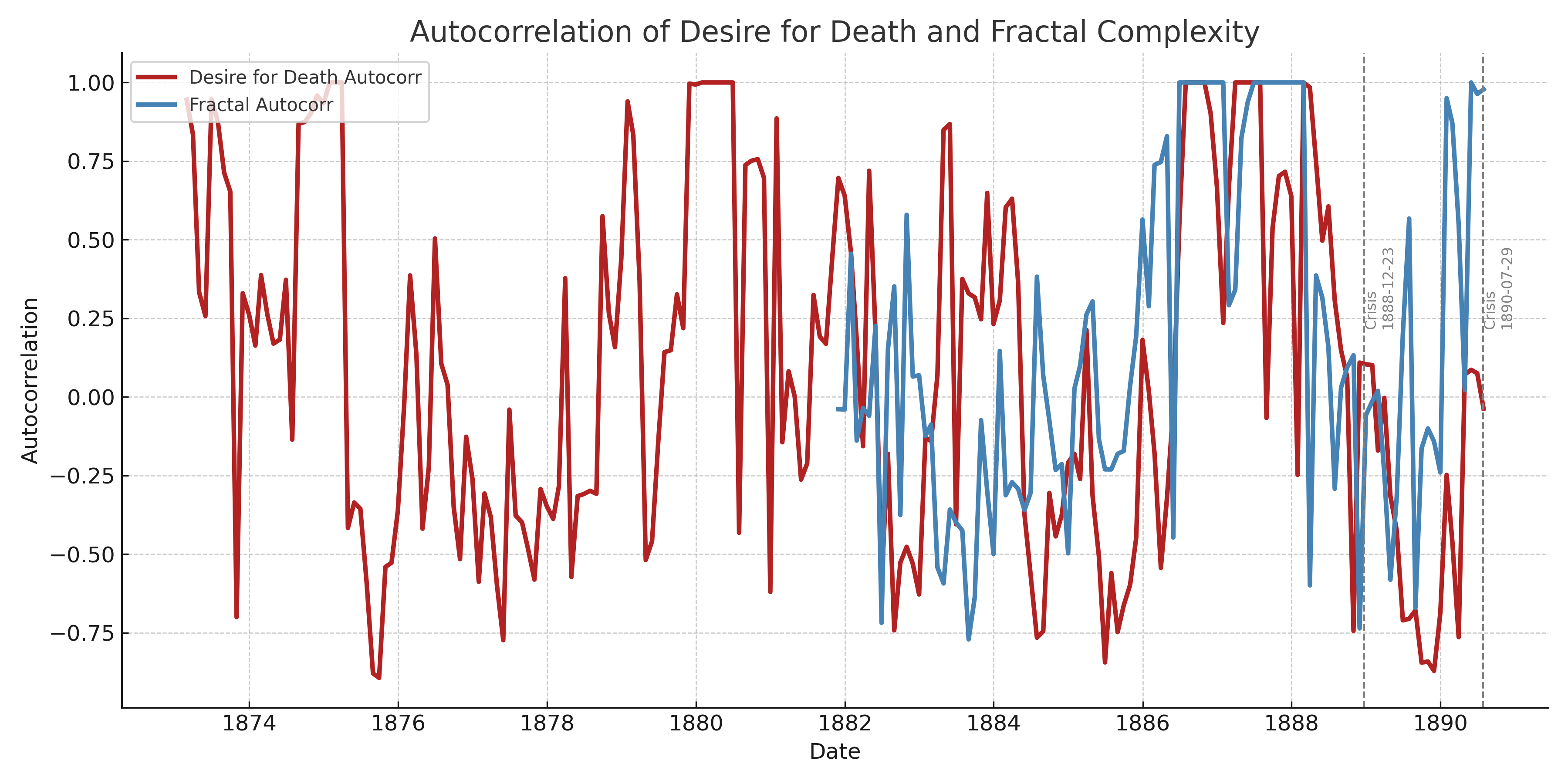}
    \caption{Autocorrelation (lag-1) plot of "Desire for Death" sentiment and the fractal dimension of paintings with a known date. There is a qualitative alignment of behavior indicating coordinated system dynamics across expression modalities despite modest linear correlation.}
    \label{fig:autocorr_original}
\end{figure}

Figure~\ref{fig:autocorr_original} provides complementary evidence by directly comparing lag-1 autocorrelation time series for desire for death sentiment (red trace) and fractal dimension (blue trace) across van Gogh's entire creative period. This visualization reveals qualitative alignment in temporal structure despite the modest linear correlation ($r = 0.193$) between the raw metric values. The 1888--1889 period shows the most dramatic coordinated autocorrelation dynamics, with both traces exhibiting sharp increases beginning in mid-1888: desire for death autocorrelation reaching $\rho \approx 0.6$ and fractal dimension autocorrelation reaching $\rho \approx 0.4$ during October--November 1888. Following the December 1888 ear incident, autocorrelation values remain elevated but show high-frequency oscillations, consistent with a system that has crossed a threshold but not yet reached a new stable equilibrium.

The qualitative alignment visible in Figure~\ref{fig:autocorr_original} demonstrates that van Gogh's psychological system exhibited coordinated slowing across fundamentally different expression channels. Textual sentiment analysis captures conscious, linguistically mediated emotional expression directed toward specific audiences (primarily Theo), while fractal dimension analysis captures fine motor skills, artistic decision-making about brushwork density and visual texture—decisions made during the physical act of painting. The fact that both channels show increased autocorrelation during the same temporal windows—particularly the 4--6 months preceding the ear incident—provides strong evidence that the observed CSD signatures reflect genuine psychological system dynamics rather than artifacts of a single measurement modality.

\begin{figure}
    \centering
    \includegraphics[width=1\linewidth]{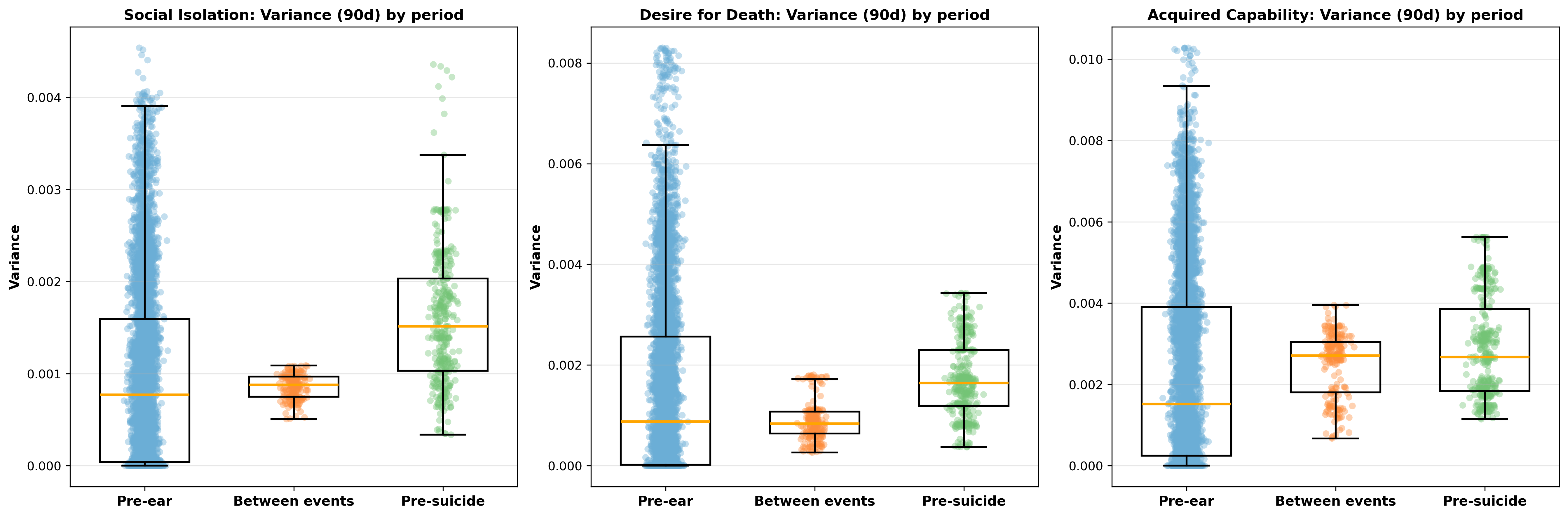}
    \caption{Box plots showing the variance in Social Isolation, Desire for Death, and Acquired Capability sentiments sorted by whether they are within 90 days of the event listed. Each data point represents the moving average of the variance with a 90 day window. }
    \label{fig:box_plot}
\end{figure}

To quantify volatility changes approaching the ear incident, we examined rolling variance estimates across three temporal periods: Pre-ear (before December 1888), Between events (ear incident to one year before suicide), and Pre-suicide (final year before July 1890). Figure~\ref{fig:box_plot} displays variance distributions for social isolation, desire for death, and acquired capability across these periods, with each data point representing the 90-day rolling variance. Though variance pre-ear is much larger owing in part to sweeping stylistic changes over the greater timespan covered (1874-1888), the pre-suicide variance is consistently higher than the between events variance despite covering comparable lengths of time.

Desire for death shows variance highest during the Pre-ear period ($\sigma^2 = 0.0088$, $\text{IQR} = 0.00020\text{--}0.0026$), decreasing substantially during Between events ($\sigma^2 = 0.00084$, $\text{IQR} = 0.0064\text{--}0.0011$), then increasing again in the Pre-suicide period ($\sigma^2 = 0.0016$, $\text{IQR} = 0.0012\text{--}0.0023$). The elevated variance before the ear incident confirms that death cognition became highly volatile during the months preceding the 1888 crisis. Social isolation demonstrates progressive increase, with median variance rising from Pre-ear ($\sigma^2 = 0.0077$, $\text{IQR} = 0.000041\text{--}0.0016$) through Between events ($\sigma^2 = 0.00088$, $\text{IQR} = 0.00076\text{--}0.00097$) to peak levels Pre-suicide ($\sigma^2 = 0.0015$, $\text{IQR} = 0.00010\text{--}0.00021$). Acquired capability exhibits the most dramatic escalation, increasing from Pre-ear ($\sigma^2 = 0.0015$, $\text{IQR} = 0.00035\text{--}0.0040$) through Between events ($\sigma^2 = 0.0027$, $\text{IQR} = 0.0018\text{--}0.0030$) to peak Pre-suicide levels ($\sigma^2 = 0.0027$, $\text{IQR} = 0.0018\text{--}0.0039$), supporting theoretical predictions about capability acquisition processes: as van Gogh experienced repeated self-harm and hospitalizations, his day-to-day capability for lethal self-injury became both elevated and increasingly erratic.

To test whether psychological and visual volatility itself varied across temporal periods, we conducted Fligner-Killeen tests comparing the variance of rolling variance estimates—essentially testing whether variance was stable or erratic within each period. All examined metrics demonstrated highly significant heterogeneity in variance-of-variance across periods (all $p < 0.0001$). For fractal dimension, variance-of-variance was highest in the Pre-suicide period ($\sigma^2_{\sigma^2} = 6.29 \times 10^{-6}$), followed by Pre-ear ($\sigma^2_{\sigma^2} = 4.64 \times 10^{-6}$), with the Between events period showing the lowest volatility ($\sigma^2_{\sigma^2} = 2.95 \times 10^{-6}$; Fligner-Killeen $H = 59.80$, $p < 0.0001$).

Sentiment metrics showed similar patterns but with more pronounced differences. The strongest effects were observed for drug/alcohol use ($H = 530.93$, $p < 0.0001$), stress ($H = 479.77$, $p < 0.0001$), and desire for death ($H = 374.95$, $p < 0.0001$). Notably, the Pre-ear period consistently exhibited the highest variance-of-variance across most psychological dimensions, suggesting increased erratic fluctuation in emotional state during the years preceding the ear incident. For instance, acquired capability showed Pre-ear variance-of-variance nearly 9-fold higher than the between events period ($\sigma^2_{\sigma^2} = 5.41 \times 10^{-6}$ vs. $6.1 \times 10^{-7}$, $H = 329.47$, $p < 0.0001$).

These findings indicate that not only did variance increase approaching critical events (as predicted by classical critical slowing down theory), but the stability of that variance itself degraded. This volatility - variance becoming more erratic rather than consistently elevated suggests the psychological system was oscillating unpredictably between states rather than smoothly transitioning, consistent with chaotic dynamics near bifurcation points. The system exhibited not just increased variance (first-order instability) but increased variance-of-variance (second-order instability), indicating a loss of regulatory capacity at multiple hierarchical levels.

The contrast between the Pre-ear and Pre-suicide patterns is particularly informative. Before the ear incident, variance-of-variance was highest across most dimensions, suggesting a system characterized by extreme erratic oscillations—lurching between states rather than following predictable trajectories. Before the suicide, variance remained elevated but became more stable (lower variance-of-variance for most sentiment dimensions except social isolation and fractal dimension), suggesting a system locked into persistent high-volatility dynamics rather than oscillating. This distinction may reflect different crisis mechanisms: the ear incident emerged from chaotic instability with unpredictable state transitions, while the suicide emerged from entrenched, persistently elevated risk with reduced capacity for recovery.

Analysis of mean psychological metric values reveals a paradoxical pattern for the ear incident: relatively low death sentiment (0.105) despite extremely high visual complexity (1.895) and elevated acquired capability (0.278), with substance use metrics elevated during this period (0.341). This dissociation between conscious death ideation and behavioral dysregulation capacity is particularly striking given the severity of the self-harm event that followed. The pattern suggests that the ear incident was not preceded by prolonged conscious suicidal contemplation (which would appear as elevated desire for death scores in letters), but rather emerged from a psychological system that had become increasingly volatile, fearless, and impulsive while conscious death cognition remained relatively suppressed.

van Gogh's letters from November--December 1888 express anxiety about Gauguin's departure and fears about his ability to sustain the collaborative artistic project, but they lack the explicit death-related content that characterizes some earlier crisis periods. The high fractal dimension during this period may reflect heightened cognitive and emotional arousal channeled into artistic production rather than linguistic expression. van Gogh was painting at an extraordinary pace during the Arles period, completing some of his most complex and energetic works during the months of psychological destabilization \citep{freedman2006yellow}. The CSD framework suggests that this intense productivity may itself have been a manifestation of system instability: the psychological system was unable to maintain equilibrium, producing both high-variance artistic output and high-variance emotional states even when conscious death cognition appeared temporarily reduced.

Beyond the ear incident, Figure~\ref{fig:historical} reveals several other periods where multiple dimensions exhibited coordinated patterns of instability, suggesting that van Gogh's psychological system experienced multiple near-threshold states throughout his final decade. The period 1881--1882 (events 1--3), encompassing van Gogh's return to Cuesmes, the hand-in-flame incident, and hospitalization for gonorrhea, shows moderate increases in volatility across desire for death and visual complexity. However, these increases are less coordinated and less sustained than those observed before the 1888 ear incident, possibly reflecting the earlier stage of van Gogh's psychological trajectory and less severe system-wide destabilization. The period 1885--1886, surrounding van Gogh's father's death and subsequent tumultuous period in Nuenen, shows coordinated increases in both desire for death and fractal dimension autocorrelation, with both series rising from near-zero or negative values to sustained positive values ($\rho \approx 0.2\text{--}0.4$ visible in Figure~\ref{fig:autocorr_original}), indicating increased temporal persistence in both psychological state and artistic complexity.

The final year of van Gogh's life (1889--1890, events 13--15 in Figure \ref{fig:historical}) shows sustained patterns of instability across multiple dimensions without subsequent relaxation, suggesting a psychological system trapped in a critical state. Unlike the ear incident, which was followed by partial psychological stabilization during the Saint-Rémy asylum period, the months following van Gogh's move to Auvers-sur-Oise (May 1890) show no evidence of the system returning to stable dynamics. The suicide attempt on July 27 occurred during a period of sustained volatility across desire for death, acquired capability, and social isolation. Figure~\ref{fig:autocorr_original} shows that in the final period, fractal dimension autocorrelation exhibits a gradual decline from $\rho \approx 0.4$ in early 1888 to negative values by mid-1890, possibly reflecting either stylistic stabilization or reduced creative output in the months before his death while hospitalized in Saint Remy. In the months preceding van Gogh's death, both the autocorrelation of Desire for Death and the fractal dimension increase from their values in 1889.

This terminal pattern - sustained instability without recovery - provides a quantitative signature distinguishing the final crisis from earlier near-threshold events. The ear incident was followed by psychological reorganization and partial resilience recovery; the suicide attempt occurred in a system that had lost the capacity to recover from perturbations. The variance analysis (Figure~\ref{fig:box_plot}) corroborates this interpretation: acquired capability variance shows progressive escalation culminating in the Pre-suicide period, while social isolation variance demonstrates continuous increase across all three periods, indicating that van Gogh's capacity for social connection was degrading throughout his final years without the temporary recoveries observed in other dimensions.

\section{Discussion}

Our results provide evidence that van Gogh's psychological trajectory exhibits mathematical signatures of critical slowing down. The coordinated variance increases across multiple dimensions before the 1888 ear incident support theoretical predictions about complex systems approaching phase transitions. This finding extends CSD theory from ecological and climate systems to individual psychological dynamics.

The multi-modal nature of CSD evidence - spanning visual complexity, death cognition, and social dimensions - suggests that psychological tipping points involve system-wide reorganization rather than isolated symptom changes. This has important implications for understanding psychological crisis dynamics and potentially developing early warning systems.

The statistically significant correlation between visual complexity and death sentiment demonstrates meaningful psychological coupling between artistic expression and conscious cognition. However, the modest correlation magnitude suggests these modalities respond to psychological stress through partially independent pathways while sharing underlying dynamical properties.

The divergent temporal patterns - decreasing visual complexity alongside slightly increasing death sentiment - reveal complex psychological system behavior approaching crisis. The social dimension patterns align closely with interpersonal theory of suicide predictions, providing convergent validation for the analytical approach.

This study demonstrates that computational approaches can successfully detect psychological transitions in historical figures through multi-modal analysis. The CSD framework provides objective criteria for identifying psychological instability periods, complementing traditional biographical analysis with quantitative validation.

\section{Conclusions}

This study demonstrates that van Gogh's psychological trajectory exhibits clear mathematical signatures of critical slowing down, particularly before the 1888 ear incident. The key contribution lies in showing that psychological transition signatures can be detected through statistical volatility patterns across multiple expression modalities.

Our findings establish that van Gogh's psychological system exhibits coordinated CSD signatures across visual and textual expression, different psychological dimensions respond through partially independent pathways to underlying stress processes, and mathematical approaches can provide objective validation for biographical accounts of psychological decline. Future directions include studying the associations among the text related to the ingestion of alcohol, absinth, or terpenes, and van Gogh's onset of bouts of illness. 

This interdisciplinary approach bridges digital humanities, computational psychology, and art history, offering replicable methods for analyzing psychological dynamics in historical figures with rich archival records. The broader implication is that psychological systems exhibit mathematical early warning signals before critical transitions, supporting the development of computational approaches to mental health research. 

\counterwithin{figure}{section}

   \section{Appendix}

\begin{figure}[H]
    \centering
    \includegraphics[width=1\linewidth]{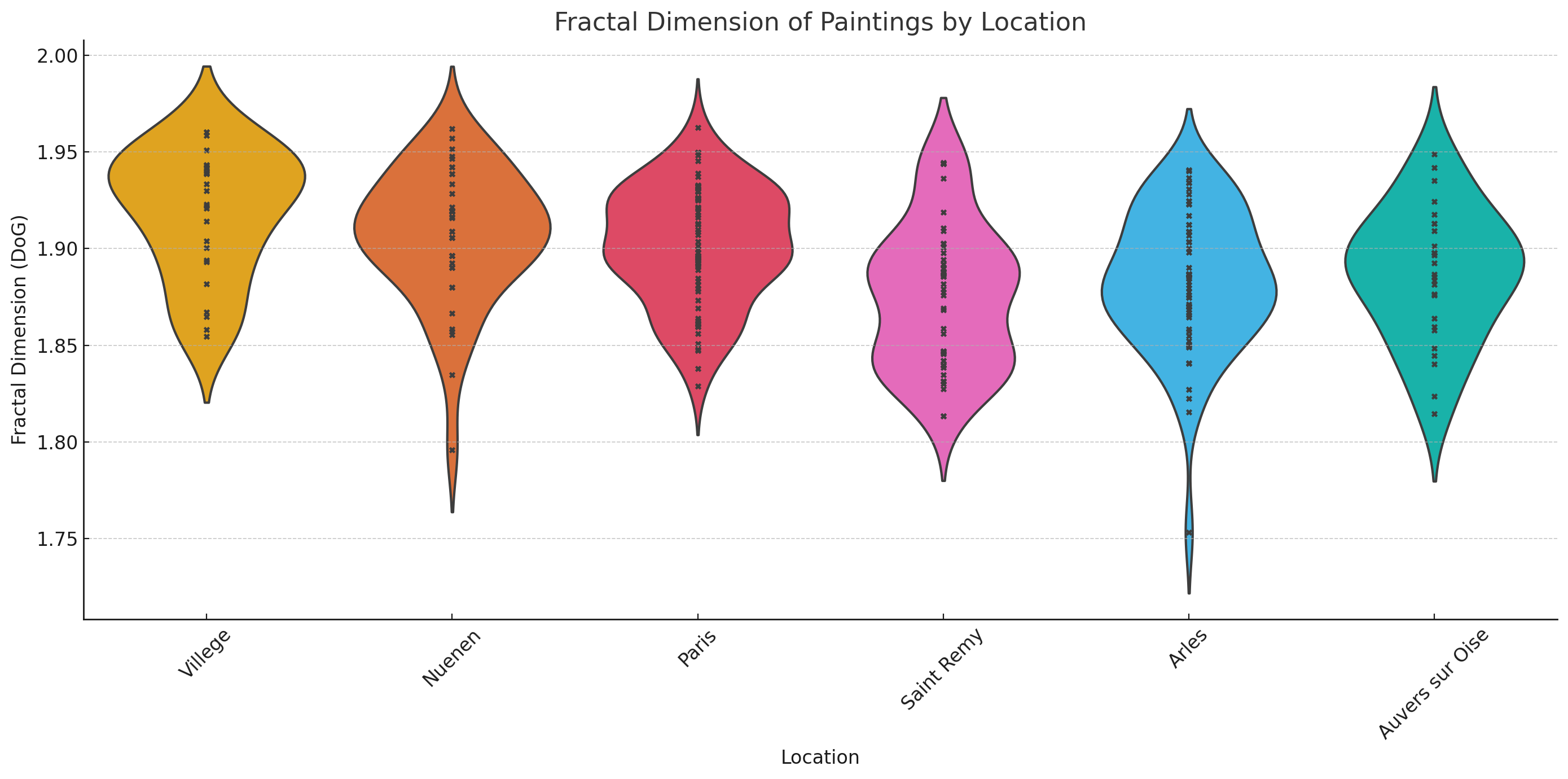}
    \caption{Violin plot of the fractal dimension of van Gogh paintings by location, showing variation in visual complexity corresponding to different life periods and psychological states.}
    \label{fig:violin_location}
\end{figure}

\begin{figure}[H]
    \centering
    \includegraphics[width=1\linewidth]{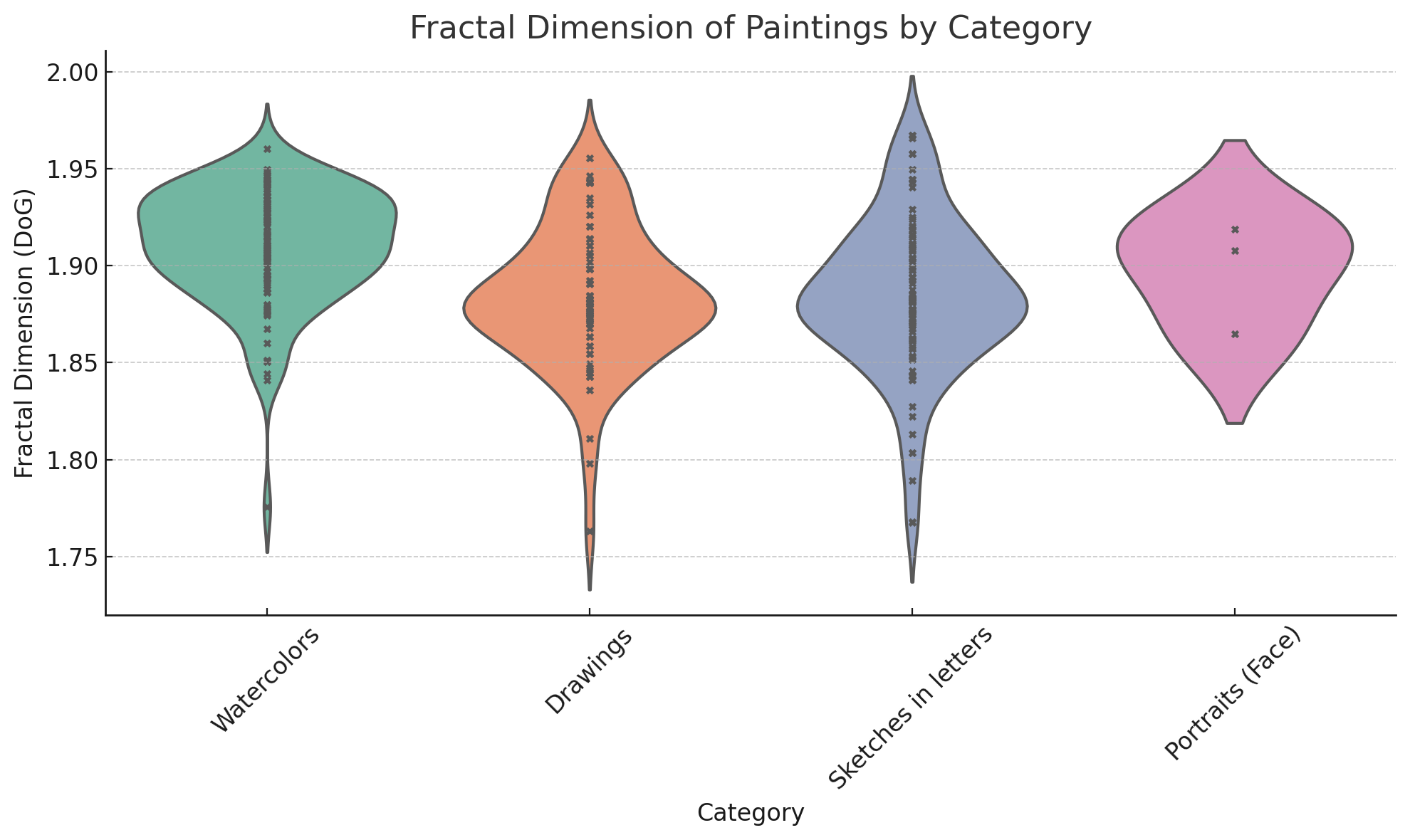}
    \caption{Violin plot of the fractal dimension of van Gogh works by type.}
    \label{fig:temporal_dist}
\end{figure}

    Figures \ref{fig:violin_location} and \ref{fig:temporal_dist} are violin plots of the fractal dimension of van Gogh's painting by location and of all his works by category. We can see small differences in fractal dimensions based on where van Gogh was living at the time as well as differences between the types of  work.
    
\section{Competing interests}
No competing interests are declared.

\section{Author contributions}
AS designed and directed the project. EG contributed to mathematical methods. MP led the neuroscience arm of the project and assisted with interpretation of results. JL contributed historical knowledge of van Gogh and qualitative expertise. All authors contributed to the preparation of the manuscript.

\section{Acknowledgments}
The authors thank the van Gogh Museum for providing access to digitized correspondence and artwork data.

\bibliography{reference.bib}

\end{document}